\documentclass[english,preprint,superscriptaddress,showpacs,prb]{revtex4}
\usepackage[T1]{fontenc}
\usepackage[latin1]{inputenc}
\usepackage{amsmath}
\usepackage{amssymb}
\usepackage{graphicx}
\usepackage{graphicx}
\usepackage{amsfonts}
\usepackage{babel}

\input{tcilatex}

\begin{document}

\author{Alexey A. Kovalev}
\affiliation{Kavli Institute of NanoScience, Delft University of Technology, 2628 CJ
Delft, The Netherlands}
\author{Gerrit E. W. Bauer}
\affiliation{Kavli Institute of NanoScience, Delft University of Technology, 2628 CJ
Delft, The Netherlands}
\author{Arne Brataas}
\affiliation{Department of Physics, Norwegian University of Science and Technology,
N-7491 Trondheim, Norway}
\title{Perpendicular spin valves with ultra-thin ferromagnetic layers}

\begin{abstract}
We address two finite size effects in perpendicular transport through
multilayers of ferromagnetic (F) and normal metal\ (N) layers: (i) the
transport properties depend on the magnetic layer thickness when of the
order or thinner than the spin-flip diffusion length and (ii) magnetic
layers with thickness approaching the magnetic coherence length become
transparent for spin currents polarized perpendicular to the magnetization.
We use magnetoelectronic circuit theory to investigate both effects on
angular magnetoresistance (aMR) and spin transfer torque in perpendicular
spin valves. We analyze recent aMR experiments to determine the spin-flip
diffusion length in the ferromagnet (Py) as well as the Py\TEXTsymbol{\vert}%
Co interface spin-mixing conductance and propose a method to measure the
ferromagnetic coherence length.
\end{abstract}

\date{\today{}}
\maketitle

\section{Introduction}

Since the discovery of the giant magnetoresistance (GMR)\cite{Baibich:prl88}
electron transport in magnetic metallic heterostructures has been studied
intensively and with considerable progress. The field developed from studies
of large area multilayers of ferromagnetic (F) and normal metals (N) in
which the current flows in the plane of the interfaces (CIP) to
nanostructures with current perpendicular to the planes (CPP).\cite%
{Gijs:ap97} Current-induced magnetization excitation has been predicted for
perpendicular F\TEXTsymbol{\vert}N\TEXTsymbol{\vert}F spin valves \cite%
{Sloncz:mmm96,Berger:prb96} and subsequently observed.\cite%
{Myers:sc99,Tsoi:prl98,Tsoi-err:prl98,Wegrowe:epl99} In these experiments
applied currents excite a spin accumulation in the normal metal spacer that
exerts a torque on the ferromagnets. When this torque overcomes the damping,
the magnetization starts to precess coherently, possibly leading to a
complete magnetization reversal.\cite{Krivorotov:sc05} By fits of the
parameters of the diffusion equation\cite{Valet:prb93} to a wealth of
experimental data of the GMR in CPP structure, the spin-dependent interface
and bulk material resistances of the most important transition metal
combinations are well known by now.\cite{Gijs:ap97,Bass:mmm99}
First-principles calculations in general agree well with the experimental
values.\cite{Galinon} Also in view of possible applications for switching
purposes in magnetic random access memories, a comparably accurate modeling
of the spin torque as a function of material combinations and applied bias
is desirable.

Physically, the spin-transfer torque is a consequence of angular momentum
conservation when a spin current polarized transverse to the magnetization
direction is absorbed at the magnetic interface.\cite{Waintal:prb00} The
transverse spin current can penetrate the ferromagnet up to a skin depth
equal to the ferromagnetic coherence length $\lambda _{c}=\pi /\left\vert
k_{\uparrow }^{F}-k_{\downarrow }^{F}\right\vert .$ In transition metals $%
\lambda _{c}$ is much smaller than all other length scales such as
spin-diffusion length or mean-free path.\cite%
{Brataas:epjb01,Sloncz:mmm02,Stiles:prb02} When the ferromagnetic layer
thickness $d_{F}\gg \lambda _{c}$ the spin-transfer torque is a pure
interface property governed by the so-called spin-mixing conductance,\cite%
{Brataas:prl00} which is accessible to first principles calculations.\cite%
{Xia:prb02} \ 

An excellent method to measure the torque and mixing conductance is the
normalized angular magnetoresistance (aMR) of perpendicular F\TEXTsymbol{%
\vert}N\TEXTsymbol{\vert}F spin valves\cite%
{Dauguet:prb96,Vedyayev:prb97,Giacomoni:02}%
\begin{equation}
\text{aMR}\left( \theta \right) =\frac{R\left( \theta \right) -R\left(
0\right) }{R\left( \pi \right) -R\left( 0\right) },
\end{equation}%
where $R\left( \theta \right) $ is electric resistance when the two
magnetizations are rotated by an angle $\theta $ with respect to each other.
Deviations of the aMR as a function of $\cos \theta $ from a straight line
are proof of a finite mixing conductance.\cite{Gerrit:prb03} Systematic new
measurements of the aMR have been carried out recently by Urazhdin \textit{%
et al.} \cite{Urazhdin:prb05} on Permalloy(Py)$|$Cu spin valves as a
function of the Py thicknesses.

Interesting effects such as non-monotonic aMR, change of sign of the
spin-transfer torque and strongly reduced critical currents for
magnetization reversal have been predicted for asymmetric spin valves.\cite%
{Kovalev:prb02,Manschot:prb04,Manschot:apl04} Asymmetry here means that the
two ferromagnets in the spin valve are not equivalent for spin transport.
This can be achieved by different thicknesses of the magnetically active
regions of otherwise identical ferromagnetic contacts, but only when the
spin-flip diffusion length in the ferromagnet $l_{sd}^{F}$ is of the order
or larger than one of the magnetic layer thicknesses. The magnetically soft
Py is the material of choice, but its spin-flip diffusion length is only $%
l_{sd}^{F}\cong 5$ nm.\cite{Bass:mmm99} Urazhdin \textit{et al.} \cite%
{Urazhdin:prb05} investigated spin valves with ultrathin $d_{F}\lesssim
l_{sd}^{F}$, which means that the analysis of these experiments requires
solution of the spin and charge diffusion equation in the ferromagnet.

Detailed calculations for transition metals\cite{Stiles:prb02,
Zwierzycki:prb05} confirm that a transverse spin current can penetrate the
ferromagnet over distances $\lesssim 1$ nm as a consequence of incomplete
destructive quantum interference. Urazhdin \textit{et al.} investigated spin
valves with Py layers of such thicknesses, claiming to observe an effect of
this transverse component on the aMR. In weak ferromagnets like CuNi or PdNi
alloys in which $\lambda _{c}$ may become larger than the scattering
mean-free path, the transverse component of spin current and accumulation
can be treated semiclassically.\cite{Shpiro:prb03} It is shown below that an
effective conductance parameter (\textquotedblleft mixing
transmission\textquotedblright ) can be introduced to parametrize transport
in both regimes.

In this paper we treat the size effects related to $d_{F}\lesssim l_{sd}^{F}$
(Section II) and $d_{F}\lesssim \lambda _{c}$ (Section III) (but $\lambda
_{c}$ much smaller than the spin diffusion length). In Section II we apply
magnetoelectronic circuit theory\cite{Brataas:prl00} combined with the
diffusion equation to the F\TEXTsymbol{\vert}N\TEXTsymbol{\vert}F\TEXTsymbol{%
\vert}N spin valves studied by Urazhdin \textit{et al..} We demonstrate that
the angular magnetoresistance provides a direct measure for the mixing
conductance\cite{Gerrit:prb03} and find that the non-monotonicity in the aMR
is indeed caused by the asymmetry as predicted. For F\TEXTsymbol{\vert}N%
\TEXTsymbol{\vert}F\TEXTsymbol{\vert}N\TEXTsymbol{\vert}F structures, that
are also of interest because of their increased spin torque,\cite%
{Berger:jap03,Nakamura:matj04} we obtain several analytical results. The
approach from Section II is generalized in Section III allowing us to treat
ultrathin ferromagnetic layers or weak ferromagnets.\cite%
{Tserkovnyak:prl02,Brataas:prl03,Zwierzycki:prb05} We find that there should
be no measurable effects of $\lambda _{c}$ on the aMR in F\TEXTsymbol{\vert}N%
\TEXTsymbol{\vert}F\TEXTsymbol{\vert}N structures, but predict that the
torque acting on the thin layer is modified. We proceed to conclude that the
coherence length should be observable in the aMR of F\TEXTsymbol{\vert}N%
\TEXTsymbol{\vert}F\TEXTsymbol{\vert}N\TEXTsymbol{\vert}F structures.
Finally, we propose a set-up to measure the ferromagnetic coherence length
in a three-terminal device.

\section{Magnetoelectronic circuit theory and diffusion equation for spin
valves}

In this Section we assume that $\lambda _{c}\ll d_{F}.$ In Part A we
recapitulate some old results: the magnetoelectronic circuit theory for spin
valves, with emphasis on the inclusion of the spin-flip diffusion in the
ferromagnetic layers when the ferromagnetic 
layer thickness $d_{F}$ is of the same order as the spin-flip
diffusion length in the ferromagnet $l_{sd}^{F}$. 
In Part B we apply these results to recent experiments by Urazhdin 
\textit{et al.} in which we can disregard spin-flip in the Cu spacers. In
Part C we present new results for symmetric F\TEXTsymbol{\vert}N\TEXTsymbol{%
\vert}F\TEXTsymbol{\vert}N\TEXTsymbol{\vert}F structures.

\subsection{Magnetoelectronic circuit theory and diffusion equation}

Magnetoelectronic circuit theory \cite{Brataas:epjb01} has been designed to
describe charge and spin transport in disordered or chaotic multi-terminal
ferromagnet-normal metal hybrid systems with non-collinear magnetizations.
The material parameters of the theory are the bulk and interface
spin-dependent conductances, as well a the so-called interface spin-mixing
conductance $G_{\uparrow \downarrow }.$ For spin valves, circuit theory can
be shown to be equivalent to a diffusion equation when $\func{Im}G_{\uparrow
\downarrow }\approx 0$, which is usually the case for intermetallic
interfaces.\cite{Kovalev:prb02} When the thickness of the ferromagnetic
metal layer $d\gg l_{sd}^{F}$, the layer bulk resistance can be effectively
replaced by that of a magnetically active region close to the interface of
thickness $l_{sd}^{F}$. When connected to a reservoir or other type of spin
sink, the effective thickness becomes $l_{sd}^{F}\tanh (d_{F}/l_{sd}^{F})$.%
\cite{Kovalev:prb02}

The aMR for general N\TEXTsymbol{\vert}F\TEXTsymbol{\vert}N\TEXTsymbol{\vert}%
F\TEXTsymbol{\vert}N structures with $\func{Im}G_{\uparrow \downarrow }=0$
as derived previously\cite{Kovalev:prb02} reads 
\begin{equation}
\Re (\theta )=R_{\uparrow \downarrow }+R_{1}+R_{2}-\frac{R_{\uparrow
\downarrow }(R_{1-}+\alpha R_{2-})^{2}+(1-\alpha
^{2})(R_{1-}^{2}R_{2}+R_{2-}^{2}(R_{1}+R_{\uparrow \downarrow }))}{%
(R_{\uparrow \downarrow }+R_{1})(R_{\uparrow \downarrow }+R_{2})-\alpha
^{2}R_{1}R_{2}}.  \label{aMR}
\end{equation}%
with $\alpha =\cos \theta ,$ $4R_{1(2)}=1/G_{1(2)\uparrow
}+1/G_{1(2)\downarrow }-2R_{\uparrow \downarrow }$, $4R_{1(2)-}=1/G_{1(2)%
\uparrow }-1/G_{1(2)\downarrow }$, $P_{1(2)}=R_{1(2)-}/R_{1(2)}$, and $%
2R_{\uparrow \downarrow }=1/G_{1\uparrow \downarrow }+1/G_{2\uparrow
\downarrow },$ where $G_{1(2)\uparrow }$ and $G_{1(2)\downarrow }$ are
conductances of the left (right) ferromagnet including the left (right)
normal layer, $G_{1\uparrow \downarrow }$ and $G_{2\uparrow \downarrow }$
are mixing conductances of the middle normal metal with adjacent ferromagnet
interfaces as shown in Fig. 1. The torques felt by first and second
ferromagnetic layer become 
\begin{equation}
\boldsymbol{\tau} _{1}/I_{0}=\frac{\hbar }{2e}\frac{1+R_{\uparrow \downarrow
}/R_{1}-\alpha P_{1}/P_{2}}{(1+R_{\uparrow \downarrow }/R_{1})(1+R_{\uparrow
\downarrow }/R_{2})-\alpha ^{2}} \mathbf{m}_{1}%
\times(\mathbf{m}_{2}\times\mathbf{m}_{1})  \label{torq1}
\end{equation}%
\begin{equation}
\boldsymbol{\tau} _{2}/I_{0}=-\frac{\hbar }{2e}\frac{1+R_{\uparrow \downarrow
}/R_{2}-\alpha P_{2}/P_{1}}{(1+R_{\uparrow \downarrow }/R_{1})(1+R_{\uparrow
\downarrow }/R_{2})-\alpha ^{2}} \mathbf{m}_{2}%
\times(\mathbf{m}_{1}\times\mathbf{m}_{2}) \label{torq2}
\end{equation}
When we approximate the mixing conductance $1/R_{\uparrow \downarrow }\,$ by
the Sharvin conductance of the normal metal, Eqs. (\ref{torq1},\ref{torq2})
coincide with the expressions in Ref. \onlinecite{Xiao:prb04} for asymmetric
N\TEXTsymbol{\vert}F\TEXTsymbol{\vert}N\TEXTsymbol{\vert}F\TEXTsymbol{\vert}%
N spin valves with $\Lambda _{L(R)}^{2}\equiv 2R_{1(2)}/R_{\uparrow
\downarrow }+1$, $P_{L(R)}\Lambda _{L(R)}^{2}=2R_{1(2)-}/R_{\uparrow
\downarrow }$.

\begin{figure}[tbp]
\caption{Definition of conductances $G_{1(2)\uparrow (\downarrow )}$ and
mixing conductances $G_{1(2)\uparrow \downarrow }$ for N\TEXTsymbol{\vert}F%
\TEXTsymbol{\vert}N\TEXTsymbol{\vert}F\TEXTsymbol{\vert}N structure.}%
\includegraphics[ 
scale=0.8]{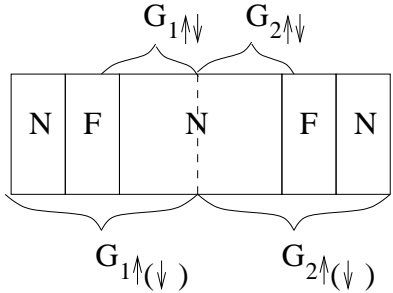}
\end{figure}

\subsection{Extraction of the mixing conductance from experiment and
asymmetric spin valves}

Most material parameters in circuit theory are those of the two-channel
resistor model. They can be determined for the collinear magnetic
configurations, \textit{i.e.} via the (binary) GMR. The only additional
parameters for the non-collinear transport are the interface mixing
conductances $G_{i\uparrow \downarrow }^{r}$, assumed here to be real. These
can be found from a single parameter fit of the experimental aMR or from
band structure calculations. A symmetric F\TEXTsymbol{\vert}N\TEXTsymbol{%
\vert}F structure is most suitable to carry out this program. The thus
obtained $G_{i\uparrow \downarrow }^{r}$ should be transferable to other
(asymmetric) structures grown by equivalent techniques. Urazhdin \textit{et
al.} fitted their experimental results for the normalized aMR by the simple
formula\cite{Giacomoni:02} that follows from circuit theory:\cite%
{Brataas:epjb01}%
\begin{equation}
\text{aMR}(\theta )=\frac{1-\cos \theta }{\chi (1+\cos \theta )+2}\,,
\label{Pratt}
\end{equation}%
For symmetric junctions we identify $\chi =2R/R_{\uparrow \downarrow }$ (see
Eq. (18) in Ref. \onlinecite{Kovalev:prb02}).

Urazhdin \emph{et al.} \cite{Urazhdin:prb05} used the structures
Nb(150)Cu(20)FeMn(8)Py($d_{1}$)Cu(10)Py($d_{2}$)Cu(20)Nb(150), where the
numbers in brackets are the thicknesses in nm. The exchange bias
antiferromagnet FeMn is treated as a perfect spin sink, which means that the
effective thickness of the left Py layer becomes $l_{sd}^{F}%
\tanh(d_{1}/l_{sd}^{F})=0.8l_{sd}$ ($d_{1}=6$ nm, $l_{sd}^{F}=5.5$ nm). Note
that this device is not exactly symmetric when $d_{2}\gg l_{sd}^{F}$ as $%
d_{1}\,\,$is$\,\,$not much larger than $l_{sd}^{F}.$ but the calculated
deviations from the fitted mixing resistances are smaller than the
experimental error bars. When we replace $d_{1}$ by $l_{sd}^{F}$ and $%
d_{2}\gg l_{sd}^{F}$ the sample is symmetric and the aMR is well represented
by Eq. (\ref{Pratt}) with $\chi=1.96$ (see Fig. 2).\cite{Urazhdin:prb05}

We can use the measured value of $\chi$ to derive the mixing conductance $%
1/(AR_{\uparrow\downarrow})$ of an interface with area $A$ by $%
R_{\uparrow\downarrow}=2R/\chi$. For comparison with first principles
calculations for point contacts based on the scattering theory of transport,%
\cite{Xia:prb02} the Sharvin resistance of the normal metal should be added%
\cite{Gerrit:prb03}%
\begin{equation*}
R_{\uparrow\downarrow}^{pc}=R_{\uparrow\downarrow}+R_{sh}
\end{equation*}
Using the notation: 
\begin{eqnarray*}
AR & = & l_{sd}\rho_{Py}^{\ast}+AR_{PyCu}^{\ast}-AR_{\uparrow\downarrow}/2 \\
AR_{-} & = & l_{sd}\rho_{Py}^{\ast}\beta_{Py}+AR_{PyCu}^{\ast}\gamma
\end{eqnarray*}
we may substitute the well established material parameters for bulk and
interface resistances of Cu\TEXTsymbol{\vert}Py \cite{Pratt:ieeem97} $%
l_{sd}\rho_{Py}^{\ast}=1.4\text{ f}\Omega\text{m}^{2}$, $l_{sd}=5.5$ nm, $%
AR_{PyCu}^{\ast}=0.5\text{ f}\Omega\text{m}^{2}$, $\beta_{Py}=0.7$, $%
\gamma=0.7$, disregarding the small bulk resistance of Cu which led us to $%
AR_{\uparrow\downarrow}=1.3\text{ f}\Omega\text{m}^{2}$ and $\eta=1.49$.
This value of the mixing resistance is larger than the Sharvin resistance $%
AR_{sh}=1/G=0.878\text{ f}\Omega\text{m}^{2}$ of Cu used by Xiao \textit{et
al.}\cite{Xiao:prb04} The point-contact mixing resistance of the Cu%
\TEXTsymbol{\vert}Py interface that should be compared with band structure
calculations is $AR_{\uparrow\downarrow}^{pc.}=2.2\text{ f}\Omega\text{m}%
^{2} $, somewhat smaller than that found in Ref. \onlinecite{Gerrit:prb03} ($%
2.56\text{ f}\Omega\text{m}^{2}$). Both results are close to the band
structure calculations \cite{Xia:prb02} of the point-contact mixing
resistance for the disordered Cu\TEXTsymbol{\vert}Co interface ($2.4\text{ f}%
\Omega\text{m}^{2}$).

In Fig. 2 we compare plots of Eq. (\ref{aMR}) with experimental aMR curves
for symmetric and asymmetric F\TEXTsymbol{\vert}N\TEXTsymbol{\vert}F%
\TEXTsymbol{\vert}N\TEXTsymbol{\vert}S multilayers,\cite{Urazhdin:cm04}
identifying the following relations between parameters: 
\begin{eqnarray*}
AR_{1} & = &
l_{sd}\rho_{Py}^{\ast}+AR_{PyCu}^{\ast}-AR_{\uparrow\downarrow}/2 \\
AR_{2} & = &
d_{2}\rho_{Py}^{\ast}+AR_{PyCu}^{\ast}+AR_{PyNb}-AR_{\uparrow\downarrow}/2 \\
AR_{1-} & = & l_{sd}\rho_{Py}^{\ast}\beta_{Py}+AR_{PyCu}^{\ast}\gamma \\
AR_{2-} & = & d_{2}\rho_{Py}^{\ast}\beta_{Py}+2AR_{PyCu}^{\ast}\gamma.
\end{eqnarray*}
We assume that the spin current into the superconductor vanishes. The
resistance between the right ferromagnet and the right reservoir was taken
to be $AR_{PyNb}=5\text{ f}\Omega\text{m}^{2}$. This is larger than the $%
AR_{PyNb}=3$ f$\Omega$m$^{2}$ reported in Ref. \onlinecite{Pratt:ieeem97},
but gives better agreement with the experiment. We observe good fits in Fig.
2 nicely reproducing the non-monotonic behavior around zero angle.

\begin{figure}[tbp]
\caption{aMR of the F\TEXTsymbol{\vert}N\TEXTsymbol{\vert}F\TEXTsymbol{\vert}%
N structure for two thicknesses of the right ferromagnetic layer $%
d_{2}=0.27l_{sd}$, $2.2l_{sd}$ ($d_{1}>l_{sd}$). The filled (large $d_{2}$)
and open (small $d_{2}$) squares are the experimental data.\protect\cite%
{Urazhdin:prb05}}\includegraphics[  scale=0.7]{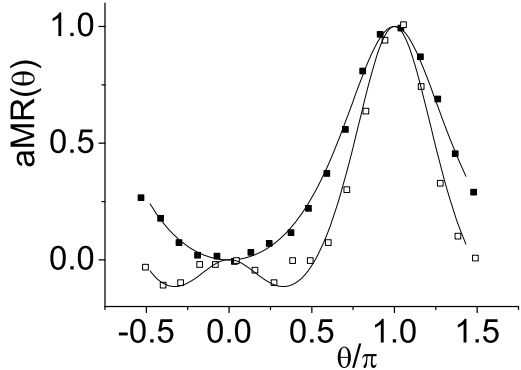}
\end{figure}
\begin{figure}[tbp]
\caption{Angular magnetoresistance $\Re (\protect\theta )-\Re (0)$ of the F%
\TEXTsymbol{\vert}N\TEXTsymbol{\vert}F\TEXTsymbol{\vert}N structure for
different thicknesses of the right ferromagnetic layer $d_{2}=0.27l_{sd}$, $%
0.5l_{sd}$, $2l_{sd}$, $2.5l_{sd}$ and $\infty $ (starting from the lower
curve respectively ($d_{1}\gg l_{sd}^{F}$, $AR_{PyNb}=3\text{ f}\Omega \text{%
m}^{2}$)). }\includegraphics[  scale=0.7]{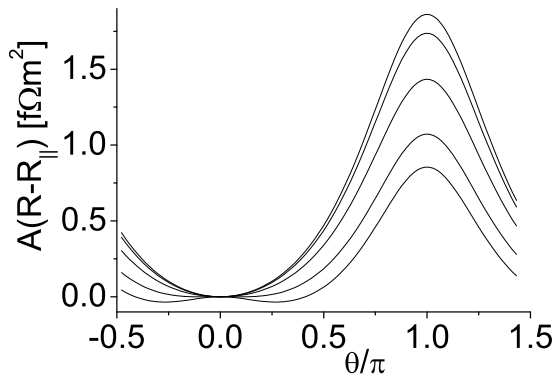}
\end{figure}
In Fig. 3 we plot the angular magnetoresistance for different thicknesses of
the right Py layer, all relative to the parallel configuration, but not
normalized to a relative scale as above. The lower curve was obtained from
Eq. (\ref{aMR}), the others were calculated numerically solving the bulk
layer spin-diffusion equation in the ferromagnet. The non-monotonic angular
magnetoresistance disappears when the right ferromagnetic layer becomes
thicker and therefore the sample more symmetric. 
\begin{figure}[tbp]
\caption{Angular magnetoresistance $\Re (\protect\theta )-\Re (0)$ and
spin-transfer torque on the left ferromagnet for the F\TEXTsymbol{\vert}N%
\TEXTsymbol{\vert}F\TEXTsymbol{\vert}N structure with right F-layer
thickness $d_{2}=0.27l_{sd}$ ($AR_{PyNb}\rightarrow \infty $, $\protect\tau %
_{0}=I_{0}\hbar /2e$). }\includegraphics[  scale=0.7]{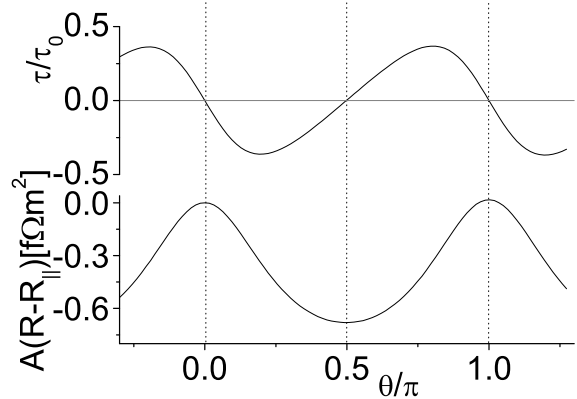}
\end{figure}
For the set of parameters in Fig. 3 the non-monotonic behavior is rather
weak but with circuit theory we can readily propose samples that maximize
the effect. The minimum of the angular magnetoresistance Eq. (\ref{aMR}) at
finite $\theta _{1}$ that coincides with a zero of the spin-transfer torque
on the left ferromagnet:\cite{Kovalev:prb02,Manschot:prb04}%
\begin{equation}
\cos \theta _{1}=\frac{(R_{\uparrow \downarrow }+R_{1})R_{2-}}{R_{1-}R_{2}}.
\label{minimum}
\end{equation}%
To observe the effect clearly, $\cos \theta _{1}$ should be small, which can
be achieved by increasing $R_{2+}$, \textit{e.g.} by the resistance of the
right-most normal metal (within the spin-flip diffusion length). In Fig. 4
we plot the angular magnetoresistance Eq. (\ref{aMR}) and the spin-transfer
torque on the left ferromagnet Eqs. (\ref{torq1}) when the resistance of the
right contact is $AR_{PyNb}\rightarrow \infty $.

\begin{figure}[tbp]
\caption{Angular magnetoresistance $\Re(\protect\theta)-\Re(0)$ of the F%
\TEXTsymbol{\vert}N\TEXTsymbol{\vert}F\TEXTsymbol{\vert}N\TEXTsymbol{\vert}F
structure for the middle F layer thicknesses $d=0.27l_{sd}$, $0.5l_{sd}$, $%
2l_{sd}$, $2.5l_{sd}$ and $\infty$ (starting from the lower curve,
respectively). The parallel resistance is subtracted.}%
\includegraphics[ 
scale=0.7]{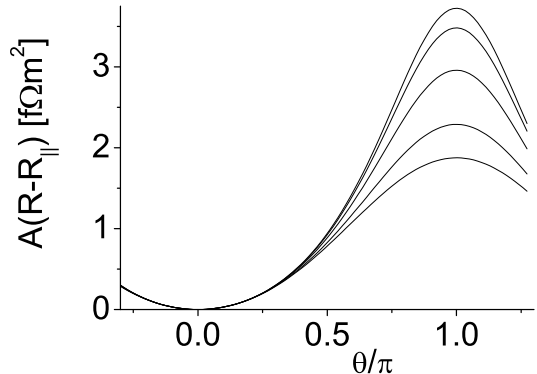}
\end{figure}

\subsection{Analysis of symmetric F\TEXTsymbol{\vert}N\TEXTsymbol{\vert}F%
\TEXTsymbol{\vert}N\TEXTsymbol{\vert}F structures}

Our approach offers analytic results for symmetric F\TEXTsymbol{\vert}N%
\TEXTsymbol{\vert}F\TEXTsymbol{\vert}N\TEXTsymbol{\vert}F structures when
the outer layers are thicker than $l_{sd}^{F}$. In Fig. 5 we plot the
angular magnetoresistance when the magnetizations of the outer layers are
kept parallel for material parameters that are the same as above and close
to set-up B from Ref. \onlinecite{Urazhdin:prb05}. When the middle layer
thickness $d_{3}\gg l_{sd}^{F}$ the angular magnetoresistance is equal to
that of two symmetric F\TEXTsymbol{\vert}N\TEXTsymbol{\vert}F structures in
series\textbf{.} The analytical formula for the angular magnetoresistance in
the regime $d_{3}\ll l_{sd}^{F}$ is presented in Appendix A. For $%
d_{3}\gtrsim 0.3l_{sd}^{F}$ we cannot disregard spin flip in the middle
layer and compute the resistances numerically.

A symmetric F\TEXTsymbol{\vert}N\TEXTsymbol{\vert}F\TEXTsymbol{\vert}N%
\TEXTsymbol{\vert}F setup with antiparallel outer layers can increase the
torque.\cite{Berger:jap03} Enhancement by a factor of $2$ was reported by S.
Nakamura\textit{\ et al.} \cite{Nakamura:matj04}. This result can be
obtained from the magnetoelectronic circuit theory (see also S. Nakamura 
\textit{et al.,} unpublished). With a current bias $I_{0}$, assuming $%
d_{3}\ll l_{sd}^{F},$ we derived a simple formula (note the similarity with
the torque on the base contact of the three-terminal spin flip transistor%
\cite{Gerrit:prb03}): 
\begin{equation}
\tau /I_{0}=\frac{\hbar }{2e}\frac{2R_{-}|\sin \theta |}{R_{\uparrow
\downarrow }+R\sin ^{2}\theta },
\end{equation}%
without invoking the parameters of the middle layer. When $d_{3}\gg l_{sd}$
we can divide system into two F\TEXTsymbol{\vert}N\TEXTsymbol{\vert}F spin
valves in series. Taking into account Eq. (3), the torque can be written
down immediately: 
\begin{eqnarray}
\tau /I_{0} &=&\tau _{FNF}(\theta )/I_{0}+\tau _{FNF}(\pi -\theta )/I_{0} \\
&=&\frac{\hbar }{2e}\frac{R_{-}|\sin \theta |}{R_{\uparrow \downarrow
}+R(1+\cos \theta )}+\frac{\hbar }{2e}\frac{R_{-}|\sin \theta |}{R_{\uparrow
\downarrow }+R(1-\cos \theta )}  \notag
\end{eqnarray}%
In Fig. 6 we plot results of these two analytic formulas as well as results
of numeric calculations for the case $d=0.8l_{sd}$. Note that these curves
are symmetric with respect to $\theta =\pi /2$. By the dashed line we plot
the torque for the corresponding symmetric F\TEXTsymbol{\vert}N\TEXTsymbol{%
\vert}F structure.

\begin{figure}[tbp]
\caption{The spin-transfer torque on the middle ferromagnet for the F%
\TEXTsymbol{\vert}N\TEXTsymbol{\vert}F\TEXTsymbol{\vert}N\TEXTsymbol{\vert}F
structure for the thickness of the middle layer $d=0.27l_{sd}$, $0.8l_{sd}$
and $10l_{sd}$ starting from the lower curve respectively (by bold line),
the same for the corresponding symmetric F\TEXTsymbol{\vert}N\TEXTsymbol{%
\vert}F structure (by dashed line), $\protect\tau_{0}=I_{0}\hbar/2e$.}%
\includegraphics[ 
scale=0.7]{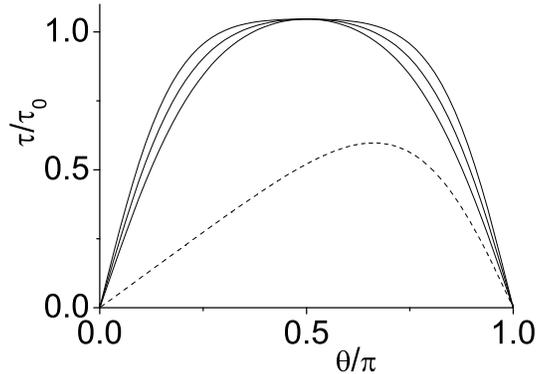}
\end{figure}

\section{Coherent regime}

Urazhdin \textit{et al}.\cite{Urazhdin:prb05}'s intentions to search for
coherence effects in ultrathin magnetic layers encouraged us to study the
regime $d_{F}\lesssim\lambda_{c}$. In this Section we formulate the
magnetoelectronic circuit theory that includes coherence effects in this
regime in two and three terminal multilayer structures. Since $\lambda_{c}$
is only a couple of monolayers, we are allowed to disregard spin-flip and
diffuse scattering in the ferromagnetic material bulk\ layers.

\subsection{Extended magnetoelectronic circuit theory}

We consider an N$1$\TEXTsymbol{\vert}F\TEXTsymbol{\vert}N$2$ circuit
element, choosing the normal metals as nodes with a possibly non-collinear
spin accumulation and the entire F layer including the interfaces as
resistive element (see Fig. 7). This allows us to treat the ferromagnet
fully quantum mechanically by scattering theory. The current through the
ferromagnet depends on the potential drop between and the spin accumulation
in each of the normal metal nodes. Spin $\mathbf{I}_{s}$ and charge $I_{0}$
currents can conveniently expressed as $2\times2$-matrices in Pauli spin
space $\widehat{I}=(\widehat{1}I_{0}+\widehat{\mathbf{\sigma}}\cdot\mathbf{I}%
_{s})/2,$ where $\widehat{\mathbf{\sigma}}$ is the vector or Pauli spin
matrices and $\widehat{1}$ the $2\times2$ unit matrix. On the normal metal
side \cite{Brataas:epjb01} in the region 2

\begin{equation}
\widehat{I}=\frac{e}{h}\{\sum_{nm}[\widehat{t^{\prime}}^{nm}\widehat{f}^{N1}(%
\widehat{t^{\prime}}^{nm})^{\dagger}-\delta_{nm} \widehat{f}^{N2}+\widehat{r}%
^{nm}\widehat{f}^{N2}(\widehat{r}^{nm})^{\dagger}]\}  \label{current}
\end{equation}
where $\widehat{r}^{mn}$ is the spin dependent reflection coefficient for
electrons reflected from channel $n$ into channel $m$ in the node 2, $%
\widehat{t^{\prime}}^{mn}$ is the spin dependent transmission coefficient
for electrons transmitted from channel $n$ in the node 1 into channel $m$ in
the node 2 and $\delta_{nm}$ is the Kronecker delta symbol.

In the absence of spin flip processes the matrices $\widehat{r}^{mn}$ and $%
\widehat{t^{\prime }}^{mn}$ should be diagonal in spin space provided the
axis $z$ is parallel to the magnetization of the ferromagnet (we are free to
chose this frame reference as it is more convenient). Expressing the
spin-dependent distribution matrices in nodes 1 and 2 via Pauli matrices; $%
\widehat{f}^{N}=\widehat{1}f_{0}^{N}+\widehat{\mathbf{\sigma }}\mathbf{f}%
_{s}^{N}$ and the unit vector \textbf{$\mathbf{m}$}$_{z}$ parallel to the
axis $z$ we obtain for spin and charge currents in the node N2: 
\begin{equation}
I_{0}=(G_{\uparrow }+G_{\downarrow })\Delta f_{0}^{N}+(G_{\uparrow
}-G_{\downarrow })\Delta \mathbf{f}_{s}^{N}\cdot \mathbf{\mathbf{m}}_{z}\,,
\label{charge}
\end{equation}%
\begin{eqnarray}
\mathbf{I}_{s} &=&\mathbf{\mathbf{m}}_{z}\left[ (G_{\uparrow }-G_{\downarrow
})\Delta f_{0}^{N}+(G_{\uparrow }+G_{\downarrow })\Delta \mathbf{f}_{s}^{N}%
\right]  \notag \\
&&-2(\mathbf{\mathbf{m}}_{z}\times \mathbf{f}_{s}^{N2}\times \mathbf{\mathbf{%
m}}_{z})\func{Re}G_{\uparrow \downarrow }^{rN2|F}+2(\mathbf{f}%
_{s}^{N2}\times \mathbf{\mathbf{m}}_{z})\func{Im}G_{\uparrow \downarrow
}^{rN2|F}  \notag \\
&&+2(\mathbf{\mathbf{m}}_{z}\times \mathbf{f}_{s}^{N1}\times \mathbf{\mathbf{%
m}}_{z})\func{Re}G_{\uparrow \downarrow }^{tN1|N2}-2(\mathbf{f}%
_{s}^{N1}\times \mathbf{\mathbf{m}}_{z})\func{Im}G_{\uparrow \downarrow
}^{tN1|N2}\,.  \label{spin}
\end{eqnarray}%
where $\Delta f_{0}^{N}=f_{0}^{N1}-f_{0}^{N2}$ and $\Delta \mathbf{f}%
_{s}^{N}=\mathbf{f}_{s}^{N1}-\mathbf{f}_{s}^{N2}$. This agrees with the
result of Ref. \onlinecite{Brataas:epjb01} except for the terms involving
the mixing transmission:\cite%
{Tserkovnyak:prl02,Brataas:prl03,Zwierzycki:prb05}%
\begin{equation*}
G_{\uparrow \downarrow }^{rN2|F}=\frac{e^{2}}{h}\sum_{nm}(\delta
_{nm}-r_{\uparrow }^{nm}\left( r_{\downarrow }^{nm}\right) ^{\ast
});\,G_{\uparrow \downarrow }^{tN1|N2}=\frac{e^{2}}{h}\sum_{nm}t_{\uparrow
}^{\prime nm}\left( t_{\downarrow }^{\prime nm}\right) ^{\ast }\,.
\end{equation*}%
The torque acting on the magnetization through the interface adjacent to N2
is the transverse component of the spin current flowing into the
ferromagnet: 
\begin{eqnarray}
\vec{\tau}_{2} &=&-2(\mathbf{\mathbf{m}}_{z}\times \mathbf{f}_{s}^{N2}\times 
\mathbf{\mathbf{m}}_{z})\func{Re}G_{\uparrow \downarrow }^{rN2|F}+2(\mathbf{f%
}_{s}^{N2}\times \mathbf{\mathbf{m}}_{z})\func{Im}G_{\uparrow \downarrow
}^{rN2|F}  \notag \\
&&+2(\mathbf{\mathbf{m}}_{z}\times \mathbf{f}_{s}^{N1}\times \mathbf{\mathbf{%
m}}_{z})\func{Re}G_{\uparrow \downarrow }^{tN1|N2}-2(\mathbf{f}%
_{s}^{N1}\times \mathbf{\mathbf{m}}_{z})\func{Im}G_{\uparrow \downarrow
}^{tN1|N2}\,.  \label{trq}
\end{eqnarray}

\begin{figure}[tbp]
\caption{A contact through a thin ferromagnet between two normal metals
nodes. The current is evaluated in the node 2.}%
\includegraphics[ 
scale=0.8]{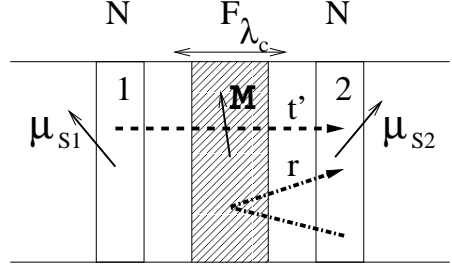}
\end{figure}
When two opposite direction of the magnetization $\mathbf{M}$ and $-\mathbf{M%
}$ are equivalent for the transport, we obtain $G_{\uparrow \downarrow
}^{tN1|N2}=G_{\uparrow \downarrow }^{tN2|N1}$ as a consequence of time
reversibility. This condition should hold in most cases (\textit{e.g.}
Stoner model is isotropic in spin space). The mixing transmission describes
the part of the transverse spin current that is not absorbed by the
ferromagnet and vanishes when the ferromagnetic layer is thicker than the
ferromagnetic coherence length $\lambda _{c}$.\cite{Zwierzycki:prb05} It is
complex, its modulus representing the transmission probability and the phase
of the rotation of the transverse spin current by the ferromagnetic exchange
field. First-principles calculations of $G_{\uparrow \downarrow }^{r}$ and $%
G_{\uparrow \downarrow }^{t}$ have been carried out by Zwierzycki \textit{et
al.}\cite{Zwierzycki:prb05} showing small variation of the first and
non-vanishing value of the second when the ferromagnetic layer becomes of
the order of several monolayers.

\subsection{Observation of ferromagnetic coherence in transport experiments}

In this section we address coherence effects due to the transmission of
transverse spin currents through ultrathin ferromagnetic layers or weak
ferromagnets. These effects should be observable in Py structures when $%
d_{F}\lesssim 1.5\,\,\text{nm.}$ Band structure calculations show that in Cu%
\TEXTsymbol{\vert}Co\TEXTsymbol{\vert}Cu structures the mixing transmission
can easily reach $G_{\uparrow \downarrow }^{t}\approx 0.1\left( G_{\uparrow
}+G_{\downarrow }\right) $ for such thicknesses.\cite{Zwierzycki:prb05}

We may draw an important conclusion from the extended magnetoelectronic
circuit theory applied to general (asymmetric) N1\TEXTsymbol{\vert}F1%
\TEXTsymbol{\vert}N2\TEXTsymbol{\vert}F2\TEXTsymbol{\vert}N3 structures:
when the nodes are chosen in the middle normal metal and in the outer normal
metals at the points that connect to the baths, a possibly finite mixing
transmission completely drops out of the charge transport equations, \textit{%
i.e.} the expressions remain exactly the same as those derived above for the
N1\TEXTsymbol{\vert}F\TEXTsymbol{\vert}N2 structure. For example, the charge
and spin currents from N1 (and similarly from N3) into N2 read 
\begin{equation}
I_{0}=(G_{\uparrow }^{\mathrm{N1|N2}}+G_{\downarrow }^{\mathrm{N1|N2}%
})\Delta \mu _{0}^{N}+(G_{\uparrow }^{\mathrm{N1|N2}}-G_{\downarrow }^{%
\mathrm{N1|N2}})\Delta \mathbf{\mu }_{s}^{N}\cdot \mathbf{\mathbf{m}_{z}}
\label{charge1}
\end{equation}%
\begin{eqnarray}
\mathbf{I}_{s} &=&\mathbf{\mathbf{m}_{z}}\left[ (G_{\uparrow }^{\mathrm{N1|N2%
}}-G_{\downarrow }^{\mathrm{N1|N2}})\Delta \mu _{0}^{N}+(G_{\uparrow }^{%
\mathrm{N1|N2}}+G_{\downarrow }^{\mathrm{N1|N2}})\Delta \mathbf{\mu }_{s}^{N}%
\right]  \notag \\
&&-2(\mathbf{\mathbf{m}_{z}}\times \mathbf{\mu }_{s}^{N2}\times \mathbf{%
\mathbf{m}_{z}})\func{Re}G_{\uparrow \downarrow }^{r\mathrm{N2|F1}}+2(%
\mathbf{\mu }_{s}^{N2}\times \mathbf{\mathbf{m}_{z}})\func{Im}G_{\uparrow
\downarrow }^{r\mathrm{N2|F1}}  \notag \\
&&+2(\mathbf{\mathbf{m}_{z}}\times \mathbf{\mu }_{s}^{N1}\times \mathbf{%
\mathbf{m}_{z}})\func{Re}G_{\uparrow \downarrow }^{t N1|N2}-2(\mathbf{\mu }%
_{s}^{N1}\times \mathbf{\mathbf{m}_{z}})\func{Im}G_{\uparrow \downarrow }^{t
N1|N2},  \label{spin1}
\end{eqnarray}%
where $\Delta \mu _{0}^{N}=\mu _{0}^{N1}-\mu _{0}^{N2}$ and $\Delta \mathbf{%
\mu }_{s}^{N}=\mathbf{\mu }_{s}^{N1}-\mathbf{\mu }_{s}^{N2}$ describe the
potential and spin accumulation drops between the left and the middle nodes.
By conservation of spin and charge currents in the center node, expression
for aMR can be derived. However, the mixing transmission does not appear in
Eqs. (\ref{charge1},\ref{spin1}) since there is no spin accumulation in the
outer nodes (reservoirs). Ferromagnets thin enough to allow transmission of
a transverse spin current can therefore not be distinguished from
conventional ones in the aMR. Our conclusions therefore disagree with the
claims of ferromagnetic coherence effects in aMR experiments on N\TEXTsymbol{%
\vert}F\TEXTsymbol{\vert}N\TEXTsymbol{\vert}F\TEXTsymbol{\vert}N structures
by Urazhdin \textit{et al}.. \cite{Urazhdin:prb05}

On the other hand, the \emph{torque} on the thin ferromagnet F2 does change: 
\begin{equation*}
\vec{\tau}_{2}=-2(\mathbf{\mathbf{m}_{z}}\times \mathbf{\mu }_{s}^{N2}\times 
\mathbf{\mathbf{m}_{z}})\func{Re}(G_{\uparrow \downarrow }^{r\mathrm{N2|F2}%
}-G_{\uparrow \downarrow }^{t\mathrm{N2|N3}})+2(\mathbf{\mu }_{s}^{N2}\times 
\mathbf{\mathbf{m}_{z}})\func{Im}(G_{\uparrow \downarrow }^{r\mathrm{N2|F2}%
}-G_{\uparrow \downarrow }^{t\mathrm{N2|N3}}).
\end{equation*}%
A parameterization of the torque via a combination $G_{\uparrow \downarrow
}^{r\mathrm{N2|F2}}-G_{\uparrow \downarrow }^{t\mathrm{N2|N3}}$ was found in
Ref. \onlinecite{Waintal:prb00} by random matrix theory, which is equivalent
with circuit theory when the number of transverse channels is large.%
\cite{Gerrit:prb03} However these authors did not discuss their
results in the limit of thin ferromagnetic layers. When $\func{Im}%
G_{\uparrow \downarrow }^{t}\approx 0$ and $\func{Im}G_{\uparrow \downarrow
}^{r}\approx 0,$ the torque $\tau _{\text{coh}}$ acting on the thin layer is
modified from the incoherent expression $\tau $ as: 
\begin{equation}
\tau _{\text{\textrm{coh}}}=\tau (G_{\uparrow \downarrow }^{r}-G_{\uparrow
\downarrow }^{t})/G_{\uparrow \downarrow }^{r}.  \label{tcoh}
\end{equation}%
Naively one may expect that the reduced absorption of the transverse spin
accumulation diminishes the torque, but this is not necessarily so (see Fig.
8). Since the mixing transmission may be negative, Eq. (\ref{tcoh}) shows
that increased torques are possible. This can be understood as follows. A
spin entering a ferromagnet will precess around an exchange field normal to
its quantization axis. A negative mixing transmission Re$G_{\uparrow
\downarrow }^{t}<0$ adds a phase factor corresponding to a rotation over an
angle $\pi $ during transmission. The outgoing spin then has a polarization
opposite to the incoming one. The magnetization torque, \textit{i.e}. the
difference between in and outgoing spin currents, consequently increases
compared to the situation in which the incoming transverse spin is absorbed
as in thick ferromagnetic layers.

In contrast to N\TEXTsymbol{\vert}F\TEXTsymbol{\vert}N\TEXTsymbol{\vert}F%
\TEXTsymbol{\vert}N structures, we find that it \emph{is} possible to
observe $G_{\uparrow\downarrow}^{t}$ in the aMR of F\TEXTsymbol{\vert}N%
\TEXTsymbol{\vert}F\TEXTsymbol{\vert}N\TEXTsymbol{\vert}F devices. We study
here the dependence of the aMR on the mixing transmission in a Py based
multilayer. In Fig. 9 we present the aMR for different mixing transmissions
in the middle layer of thickness $d_{F}=0.27l_{sd}$. Unfortunately, it seems
difficult to obtain quantitative values for the mixing transmission from
experiments since the dependence of the aMR on $G_{\uparrow\downarrow}^{t}$
is rather weak.

When the coherence length becomes larger than the scattering mean-free path,
which can occur in weak ferromagnets like PdNi or CuNi, the transverse spin
accumulation should be treated by a diffusion equation.\cite{Zhang:prl02}
The result can be parametrized again in terms of a mixing transmission,
which can subsequently be used in our circuit theory.

\begin{figure}[tbp]
\caption{The torque on the thin right layer of thickness $d=0.27l_{sd}$ for F%
\TEXTsymbol{\vert}N\TEXTsymbol{\vert}F structure. The left layer has
thickness $d\gg l_{sd}^{F}$. The curves starting from the lower one
respectively, $\text{Re}(1/G_{\uparrow\downarrow}^{t})=5\text{f}\Omega\text{m%
}^{2}$, $\text{Im}(1/G_{\uparrow\downarrow}^{t})=\infty$; $\text{Re}%
(1/G_{\uparrow\downarrow}^{t})=\infty$, $\text{Im}(1/G_{\uparrow\downarrow
t}^{t})=5\text{f}\Omega\text{m}^{2}$; $\text{Re}(1/G_{\uparrow%
\downarrow}^{t})=\infty$, $\text{Im}(1/G_{\uparrow\downarrow}^{t})=\infty$; $%
\text{Re}(1/G_{\uparrow\downarrow}^{t})=-5\text{f}\Omega\text{m}^{2}$, $%
\text{Im}(1/G_{\uparrow\downarrow}^{t})=\infty$ ($\protect\tau%
_{0}=I_{0}\hbar/2e$). }\includegraphics[
scale=0.6]{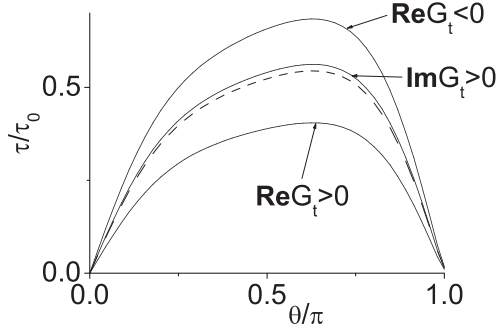}
\end{figure}

\begin{figure}[tbp]
\caption{aMR of F\TEXTsymbol{\vert}N\TEXTsymbol{\vert}F\TEXTsymbol{\vert}N%
\TEXTsymbol{\vert}F structure for the thickness of the middle layer $%
d=0.27l_{sd}$. Outer layers are antiparallel with $d\gg l_{sd}^{F}$. The
curves starting from the lower one respectively, $\text{Re}%
(1/G_{\uparrow\downarrow}^{t})=-5\text{f}\Omega\text{m}^{2}$, $\text{Im}%
(1/G_{\uparrow\downarrow}^{t})=\infty$; $\text{Re}(1/G_{\uparrow%
\downarrow}^{t})=\infty$, $\text{Im}(1/G_{\uparrow\downarrow t}^{t})=5\text{f%
}\Omega\text{m}^{2}$; $\text{Re}(1/G_{\uparrow\downarrow}^{t})=\infty$, $%
\text{Im}(1/G_{\uparrow\downarrow}^{t})=\infty$; $\text{Re}%
(1/G_{\uparrow\downarrow}^{t})=5\text{f}\Omega\text{m}^{2}$, $\text{Im}%
(1/G_{\uparrow\downarrow}^{t})=\infty$. }%
\includegraphics[
scale=0.6]{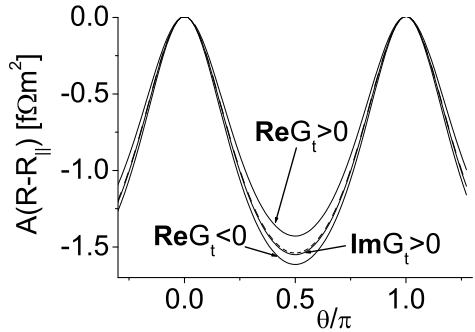}
\end{figure}

\subsection{Three terminal device for observation of coherence effects}

Finally, we propose an experiment that should be more sensitive to
ferromagnetic coherence. We suggest the setup shown in Fig. 10 that is
analogous to the spin-torque transistor\cite{Bauer:apl03} and the
magnetoelectronic spin-echo\cite{Brataas:prl03} concepts. A current through
the antiparallel ferromagnets F1 and F2 excites a spin accumulation in the
normal metal N1. This spin accumulation can transmit F3 only when its
thickness is less than $\lambda _{c}$. In that case a spin accumulation is
induced in the upper normal metal N2 that can be detected as a voltage
depending on the magnetization angle $\theta $ of the analyzing ferromagnet
F4. We assume here that N1 is smaller than its spin-flip diffusion length
(Cu is a good candidate with spin-diffusion lengths of up to a micron) such
that the spin accumulation is constant under the contact to F3. Otherwise
the signal at the ferromagnet F4 is diminished since part of the spin
accumulation in N1 is lost due to spin-flip processes.

When the $G_{\uparrow\downarrow}^{t}$ of F3 is smaller than its $%
G_{\uparrow\downarrow}^{r}$ and ferromagnet F4 is not too leaky for the spin
current (\textit{e.g}\emph{.} connected via a tunnel junction) the spin
accumulation in N2 can be found from Eqs. (\ref{charge},\ref{spin}) in terms
of the spin accumulation in N1:%
\begin{eqnarray}
\mathbf{\mu}_{S2} & = & \frac{\left|\mu_{S1}\right|}{(\func{Re}%
G_{\uparrow\downarrow}^{r})^{2}+(\func{Im}G_{\uparrow\downarrow}^{r})^{2}}%
\left(%
\begin{array}{c}
0 \\ 
\func{Re}G_{\uparrow\downarrow}^{r}\func{Re}G_{\uparrow\downarrow}^{t}+\func{%
Im}G_{\uparrow\downarrow}^{r}\func{Im}G_{\uparrow\downarrow}^{t} \\ 
\func{Re}G_{\uparrow\downarrow}^{r}\func{Im}G_{\uparrow\downarrow}^{t}-\func{%
Im}G_{\uparrow\downarrow}^{r}\func{Re}G_{\uparrow\downarrow}^{t}%
\end{array}%
\right)  \label{SA} \\
& & \overset{\func{Im}G_{\uparrow\downarrow}^{r}\rightarrow0}{\approx}\frac{%
\left|\mu_{S1}\right|}{\func{Re}G_{\uparrow\downarrow}^{r}}\left(%
\begin{array}{c}
0 \\ 
\func{Re}G_{\uparrow\downarrow}^{t} \\ 
\func{Im}G_{\uparrow\downarrow}^{t}%
\end{array}%
\right)  \label{SA2}
\end{eqnarray}
where Eq. (\ref{SA2}) holds to a good approximation when the layer F3 is
metallic. The spin accumulation is indeed coherently rotated by the exchange
field in F3. The angle dependence of the potential in F4 is $%
U(\theta)\approx\mu_{S1}P|G_{\uparrow\downarrow}^{t}|\cos\theta/\left(\func{%
Re}G_{\uparrow\downarrow}^{r}\right)$ with maximum along $\mathbf{\mu}_{S2}$%
, where $P$ is the polarization of the contact N2\TEXTsymbol{\vert}F4.

When the $G_{\uparrow\downarrow}^{t}$ of F3 is not smaller than $%
G_{\uparrow\downarrow}^{r}$ (or the spin current leak into F4 is
significant), the spin accumulation $\mathbf{\mu}_{S1}$ is affected by $%
\mathbf{\mu}_{S2}$ and the final expressions are more complicated.

\begin{figure}[tbp]
\caption{An experimental setup to observe the mixing transmission and
measure the ferromagnetic coherence length in ferromagnet F3. The spin
accumulation $\vec{\protect\mu}_{S2}$ in the normal metal N2 is measured via
the angular dependence of the potential $U(\protect\theta)$ of the
ferromagnet F4 that is weakly coupled to N2.}%
\includegraphics[ 
scale=0.7]{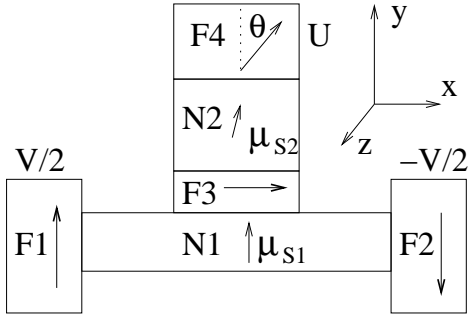}
\end{figure}
An angle dependence of $U(\theta)$ provides a direct proof of a finite
mixing transmission. The ferromagnetic coherence length can be determined by
repeating experiments for a number of layer thicknesses of F3. Such a direct
experimental evidence should help to get a grip on this important parameter $%
\lambda_{c}$.

\section{Conclusion}

In this paper we extracted the spin-mixing conductance of a Py$|$Cu
interface from the experimental data of Urazhdin \textit{et al.} using
material parameters measured independently by the MSU collaboration. We find
good agreement with experiments on asymmetric F\TEXTsymbol{\vert}N%
\TEXTsymbol{\vert}F\TEXTsymbol{\vert}N multilayers, reproducing
quantitatively the non-monotonic aMR that we predicted earlier.\cite%
{Kovalev:prb02,Manschot:prb04,Manschot:apl04} Magnetoelectronic circuit
theory together with the diffusion equation is a convenient tool for the
data analysis when the spin-flip diffusion length in the ferromagnet is of
the same order as the layer thickness. We suggest carrying out
current-induced magnetization reversal experiments on samples that display
the non-monotonic aMR since we predict anomalous magnetization trajectories
due to a vanishing torque at finite magnetization angle.\cite%
{Kovalev:prb02,Manschot:prb04,Manschot:apl04} We also study the effects of
the finite ferromagnetic coherence length in ultrathin ferromagnetic films
or weak ferromagnets. For this purpose the magnetoelectronic circuit theory
is extended to treat phase coherent transport in the ferromagnet. A
coherence length that is larger than the ferromagnetic layer thickness does
not modify the aMR of N\TEXTsymbol{\vert}F\TEXTsymbol{\vert}N\TEXTsymbol{%
\vert}F\TEXTsymbol{\vert}N structures, but a small effect should exist in F%
\TEXTsymbol{\vert}N\TEXTsymbol{\vert}F\TEXTsymbol{\vert}N\TEXTsymbol{\vert}F
structures. In contrast, the spin-transfer torque is affected more strongly
and may even be increased by the spin-coherence when the exchange field
rotates the transverse spin current polarization by the angle $\pi $.
Finally, we propose a three-terminal device that should allow experimental
determination of the ferromagnetic coherence length.

\appendix

\section{Analytical results for F|N|F|N|F structure.}

The aMR of a F$(\uparrow)$\TEXTsymbol{\vert}N\TEXTsymbol{\vert}F$(\theta)$%
\TEXTsymbol{\vert}N\TEXTsymbol{\vert}F$(\uparrow/\downarrow)$ CPP pillar can
be described analytically when the thick outer layers are parallel or
antiparallel, respectively.%
\begin{eqnarray}
R(\theta) & = & 2(R_{\uparrow\downarrow}+R)+R_{M}  \notag \\
& & -\frac{R_{\uparrow\downarrow}(R_{M-}^{2}+4R_{-}(R_{-}+R_{M-}%
\alpha))+(2R_{M}R_{-}^{2}+R_{M-}^{2}R)(1-\alpha^{2})}{(R_{\uparrow%
\downarrow}+R)(2R_{\uparrow\downarrow}+R_{M})-RR_{M}\alpha^{2}},
\end{eqnarray}
\begin{eqnarray}
R(\theta) & = & 2(R_{\uparrow\downarrow}+R)+R_{M}-\frac{2R_{-}^{2}(1-%
\alpha^{2})}{R_{\uparrow\downarrow}+R(1-\alpha^{2})}  \notag \\
& & -\frac{2R_{M-}^{2}(R_{\uparrow\downarrow}+R(1-\alpha^{2}))}{%
(R_{\uparrow\downarrow}+R)(2R_{\uparrow\downarrow}+R_{M})-RR_{M}\alpha^{2}},
\end{eqnarray}
where $\alpha=\cos\theta$, $4R+2R_{\uparrow\downarrow}=\frac{1}{G_{\uparrow}}%
+\frac{1}{G_{\downarrow}}$ , $4R_{-}=\frac{1}{G_{\uparrow}}-\frac{1}{%
G_{\downarrow}}$ for the outer layers. The mixing resistance for two
interfaces adjacent to any normal metal $R_{\uparrow\downarrow}=\frac{1}{%
G_{\uparrow\downarrow}^{r}}$ (we assume all interfaces identical). For the
middle layer $4R_{M}=\frac{1}{G_{\uparrow}}+\frac{1}{G_{\downarrow}}%
-4R_{\uparrow\downarrow}$ , $4R_{M-}=\frac{1}{G_{\uparrow}}-\frac{1}{%
G_{\downarrow}}$.

\bibliographystyle{apsrev}

\begin{thebibliography}{37}
\expandafter\ifx\csname natexlab\endcsname\relax\def\natexlab#1{#1}\fi
\expandafter\ifx\csname bibnamefont\endcsname\relax
  \def\bibnamefont#1{#1}\fi
\expandafter\ifx\csname bibfnamefont\endcsname\relax
  \def\bibfnamefont#1{#1}\fi
\expandafter\ifx\csname citenamefont\endcsname\relax
  \def\citenamefont#1{#1}\fi
\expandafter\ifx\csname url\endcsname\relax
  \def\url#1{\texttt{#1}}\fi
\expandafter\ifx\csname urlprefix\endcsname\relax\def\urlprefix{URL }\fi
\providecommand{\bibinfo}[2]{#2}
\providecommand{\eprint}[2][]{\url{#2}}

\bibitem[{\citenamefont{Baibich et~al.}(1988)\citenamefont{Baibich, Broto,
  Fert, Dau, Petroff, Eitenne, Creuzet, Friederich, and
  Chazelas}}]{Baibich:prl88}
\bibinfo{author}{\bibfnamefont{M.}~\bibnamefont{Baibich}},
  \bibinfo{author}{\bibfnamefont{J.~M.} \bibnamefont{Broto}},
  \bibinfo{author}{\bibfnamefont{A.}~\bibnamefont{Fert}},
  \bibinfo{author}{\bibfnamefont{F.~N.~V.} \bibnamefont{Dau}},
  \bibinfo{author}{\bibfnamefont{F.}~\bibnamefont{Petroff}},
  \bibinfo{author}{\bibfnamefont{P.}~\bibnamefont{Eitenne}},
  \bibinfo{author}{\bibfnamefont{G.}~\bibnamefont{Creuzet}},
  \bibinfo{author}{\bibfnamefont{A.}~\bibnamefont{Friederich}},
  \bibnamefont{and} \bibinfo{author}{\bibfnamefont{J.}~\bibnamefont{Chazelas}},
  \bibinfo{journal}{Phys. Rev. Lett.} \textbf{\bibinfo{volume}{61}},
  \bibinfo{pages}{2472} (\bibinfo{year}{1988}).

\bibitem[{\citenamefont{Gijs and Bauer}(1997)}]{Gijs:ap97}
\bibinfo{author}{\bibfnamefont{M.~A.~M.} \bibnamefont{Gijs}} \bibnamefont{and}
  \bibinfo{author}{\bibfnamefont{G.~E.~W.} \bibnamefont{Bauer}},
  \bibinfo{journal}{Adv. Phys.} \textbf{\bibinfo{volume}{46}},
  \bibinfo{pages}{285} (\bibinfo{year}{1997}).

\bibitem[{\citenamefont{Slonczewski}(1996)}]{Sloncz:mmm96}
\bibinfo{author}{\bibfnamefont{J.~C.} \bibnamefont{Slonczewski}},
  \bibinfo{journal}{J. Magn. Magn. Mater.} \textbf{\bibinfo{volume}{159}},
  \bibinfo{pages}{L1} (\bibinfo{year}{1996}).

\bibitem[{\citenamefont{Berger}(1996)}]{Berger:prb96}
\bibinfo{author}{\bibfnamefont{L.}~\bibnamefont{Berger}},
  \bibinfo{journal}{Phys. Rev. B} \textbf{\bibinfo{volume}{54}},
  \bibinfo{pages}{9353} (\bibinfo{year}{1996}).

\bibitem[{\citenamefont{Myers et~al.}(1999)\citenamefont{Myers, Ralph, Katine,
  Louie, and Buhrman}}]{Myers:sc99}
\bibinfo{author}{\bibfnamefont{E.~B.} \bibnamefont{Myers}},
  \bibinfo{author}{\bibfnamefont{D.~C.} \bibnamefont{Ralph}},
  \bibinfo{author}{\bibfnamefont{J.~A.} \bibnamefont{Katine}},
  \bibinfo{author}{\bibfnamefont{R.~N.} \bibnamefont{Louie}}, \bibnamefont{and}
  \bibinfo{author}{\bibfnamefont{R.~A.} \bibnamefont{Buhrman}},
  \bibinfo{journal}{Science} \textbf{\bibinfo{volume}{285}},
  \bibinfo{pages}{867} (\bibinfo{year}{1999}).

\bibitem[{\citenamefont{Tsoi et~al.}(1998{\natexlab{a}})\citenamefont{Tsoi,
  Jansen, Bass, Chiang, Seck, Tsoi, and Wyder}}]{Tsoi:prl98}
\bibinfo{author}{\bibfnamefont{M.}~\bibnamefont{Tsoi}},
  \bibinfo{author}{\bibfnamefont{A.}~\bibnamefont{Jansen}},
  \bibinfo{author}{\bibfnamefont{J.}~\bibnamefont{Bass}},
  \bibinfo{author}{\bibfnamefont{W.}~\bibnamefont{Chiang}},
  \bibinfo{author}{\bibfnamefont{M.}~\bibnamefont{Seck}},
  \bibinfo{author}{\bibfnamefont{V.}~\bibnamefont{Tsoi}}, \bibnamefont{and}
  \bibinfo{author}{\bibfnamefont{P.}~\bibnamefont{Wyder}},
  \bibinfo{journal}{Phys. Rev. Lett.} \textbf{\bibinfo{volume}{80}},
  \bibinfo{pages}{4281} (\bibinfo{year}{1998}{\natexlab{a}}).

\bibitem[{\citenamefont{Tsoi et~al.}(1998{\natexlab{b}})\citenamefont{Tsoi,
  Jansen, Bass, Chiang, Seck, Tsoi, and Wyder}}]{Tsoi-err:prl98}
\bibinfo{author}{\bibfnamefont{M.}~\bibnamefont{Tsoi}},
  \bibinfo{author}{\bibfnamefont{A.}~\bibnamefont{Jansen}},
  \bibinfo{author}{\bibfnamefont{J.}~\bibnamefont{Bass}},
  \bibinfo{author}{\bibfnamefont{W.}~\bibnamefont{Chiang}},
  \bibinfo{author}{\bibfnamefont{M.}~\bibnamefont{Seck}},
  \bibinfo{author}{\bibfnamefont{V.}~\bibnamefont{Tsoi}}, \bibnamefont{and}
  \bibinfo{author}{\bibfnamefont{P.}~\bibnamefont{Wyder}},
  \bibinfo{journal}{Phys. Rev. Lett.} \textbf{\bibinfo{volume}{81}},
  \bibinfo{pages}{493} (\bibinfo{year}{1998}{\natexlab{b}}).

\bibitem[{\citenamefont{Wegrowe et~al.}(1999)\citenamefont{Wegrowe, Kelly,
  Jaccard, Guittienne, and Ansermet}}]{Wegrowe:epl99}
\bibinfo{author}{\bibfnamefont{J.}~\bibnamefont{Wegrowe}},
  \bibinfo{author}{\bibfnamefont{D.}~\bibnamefont{Kelly}},
  \bibinfo{author}{\bibfnamefont{Y.}~\bibnamefont{Jaccard}},
  \bibinfo{author}{\bibfnamefont{P.}~\bibnamefont{Guittienne}},
  \bibnamefont{and} \bibinfo{author}{\bibfnamefont{J.}~\bibnamefont{Ansermet}},
  \bibinfo{journal}{Europhys. Lett.} \textbf{\bibinfo{volume}{45}},
  \bibinfo{pages}{626} (\bibinfo{year}{1999}).

\bibitem[{\citenamefont{Krivorotov et~al.}(2005)\citenamefont{Krivorotov,
  Emley, Sankey, Kiselev, Ralph, and Buhrman}}]{Krivorotov:sc05}
\bibinfo{author}{\bibfnamefont{I.~N.} \bibnamefont{Krivorotov}},
  \bibinfo{author}{\bibfnamefont{N.~C.} \bibnamefont{Emley}},
  \bibinfo{author}{\bibfnamefont{J.~C.} \bibnamefont{Sankey}},
  \bibinfo{author}{\bibfnamefont{S.~I.} \bibnamefont{Kiselev}},
  \bibinfo{author}{\bibfnamefont{D.~C.} \bibnamefont{Ralph}}, \bibnamefont{and}
  \bibinfo{author}{\bibfnamefont{R.~A.} \bibnamefont{Buhrman}},
  \bibinfo{journal}{Science} \textbf{\bibinfo{volume}{307}},
  \bibinfo{pages}{228} (\bibinfo{year}{2005}).

\bibitem[{\citenamefont{Valet and Fert}(1993)}]{Valet:prb93}
\bibinfo{author}{\bibfnamefont{T.}~\bibnamefont{Valet}} \bibnamefont{and}
  \bibinfo{author}{\bibfnamefont{A.}~\bibnamefont{Fert}},
  \bibinfo{journal}{Phys. Rev. B} \textbf{\bibinfo{volume}{48}},
  \bibinfo{pages}{7099} (\bibinfo{year}{1993}).

\bibitem[{\citenamefont{Bass and {Pratt, Jr.}}(1999)}]{Bass:mmm99}
\bibinfo{author}{\bibfnamefont{J.}~\bibnamefont{Bass}} \bibnamefont{and}
  \bibinfo{author}{\bibfnamefont{W.}~\bibnamefont{{Pratt, Jr.}}},
  \bibinfo{journal}{J. Magn. Magn. Mater.} \textbf{\bibinfo{volume}{200}},
  \bibinfo{pages}{274} (\bibinfo{year}{1999}).

\bibitem[{\citenamefont{Galinon et~al.}(preprint)\citenamefont{Galinon,
  Tewolde, Loloee, Chiang, Olson, Kurt, {Pratt, Jr.}, Bass, Xu, Xia
  et~al.}}]{Galinon}
\bibinfo{author}{\bibfnamefont{C.}~\bibnamefont{Galinon}},
  \bibinfo{author}{\bibfnamefont{K.}~\bibnamefont{Tewolde}},
  \bibinfo{author}{\bibfnamefont{R.}~\bibnamefont{Loloee}},
  \bibinfo{author}{\bibfnamefont{W.-C.} \bibnamefont{Chiang}},
  \bibinfo{author}{\bibfnamefont{S.}~\bibnamefont{Olson}},
  \bibinfo{author}{\bibfnamefont{H.}~\bibnamefont{Kurt}},
  \bibinfo{author}{\bibfnamefont{W.}~\bibnamefont{{Pratt, Jr.}}},
  \bibinfo{author}{\bibfnamefont{J.}~\bibnamefont{Bass}},
  \bibinfo{author}{\bibfnamefont{P.}~\bibnamefont{Xu}},
  \bibinfo{author}{\bibfnamefont{K.}~\bibnamefont{Xia}}, \bibnamefont{et~al.}
  (\bibinfo{year}{preprint}).

\bibitem[{\citenamefont{Waintal et~al.}(2000)\citenamefont{Waintal, Myers,
  Brouwer, and Ralph}}]{Waintal:prb00}
\bibinfo{author}{\bibfnamefont{X.}~\bibnamefont{Waintal}},
  \bibinfo{author}{\bibfnamefont{E.~B.} \bibnamefont{Myers}},
  \bibinfo{author}{\bibfnamefont{P.~W.} \bibnamefont{Brouwer}},
  \bibnamefont{and} \bibinfo{author}{\bibfnamefont{D.~C.} \bibnamefont{Ralph}},
  \bibinfo{journal}{Phys. Rev. B} \textbf{\bibinfo{volume}{62}},
  \bibinfo{pages}{12317} (\bibinfo{year}{2000}).

\bibitem[{\citenamefont{Brataas et~al.}(2001)\citenamefont{Brataas, Nazarov,
  and Bauer}}]{Brataas:epjb01}
\bibinfo{author}{\bibfnamefont{A.}~\bibnamefont{Brataas}},
  \bibinfo{author}{\bibfnamefont{Y.~V.} \bibnamefont{Nazarov}},
  \bibnamefont{and} \bibinfo{author}{\bibfnamefont{G.~E.~W.}
  \bibnamefont{Bauer}}, \bibinfo{journal}{Eur. Phys. J. B}
  \textbf{\bibinfo{volume}{22}}, \bibinfo{pages}{99} (\bibinfo{year}{2001}).

\bibitem[{\citenamefont{Slonczewski}(2002)}]{Sloncz:mmm02}
\bibinfo{author}{\bibfnamefont{J.~C.} \bibnamefont{Slonczewski}},
  \bibinfo{journal}{J. Magn. Magn. Mater.} \textbf{\bibinfo{volume}{247}},
  \bibinfo{pages}{324} (\bibinfo{year}{2002}).

\bibitem[{\citenamefont{Stiles and Zangwill}(2002)}]{Stiles:prb02}
\bibinfo{author}{\bibfnamefont{M.~D.} \bibnamefont{Stiles}} \bibnamefont{and}
  \bibinfo{author}{\bibfnamefont{A.}~\bibnamefont{Zangwill}},
  \bibinfo{journal}{Phys. Rev. B} \textbf{\bibinfo{volume}{66}},
  \bibinfo{pages}{014407} (\bibinfo{year}{2002}).

\bibitem[{\citenamefont{Brataas et~al.}(2000)\citenamefont{Brataas, Nazarov,
  and Bauer}}]{Brataas:prl00}
\bibinfo{author}{\bibfnamefont{A.}~\bibnamefont{Brataas}},
  \bibinfo{author}{\bibfnamefont{Y.~V.} \bibnamefont{Nazarov}},
  \bibnamefont{and} \bibinfo{author}{\bibfnamefont{G.~E.~W.}
  \bibnamefont{Bauer}}, \bibinfo{journal}{Phys. Rev. Lett.}
  \textbf{\bibinfo{volume}{84}}, \bibinfo{pages}{2481} (\bibinfo{year}{2000}).

\bibitem[{\citenamefont{Xia et~al.}(2002)\citenamefont{Xia, Kelly, Bauer,
  Brataas, and Turek}}]{Xia:prb02}
\bibinfo{author}{\bibfnamefont{K.}~\bibnamefont{Xia}},
  \bibinfo{author}{\bibfnamefont{P.~J.} \bibnamefont{Kelly}},
  \bibinfo{author}{\bibfnamefont{G.~E.~W.} \bibnamefont{Bauer}},
  \bibinfo{author}{\bibfnamefont{A.}~\bibnamefont{Brataas}}, \bibnamefont{and}
  \bibinfo{author}{\bibfnamefont{I.}~\bibnamefont{Turek}},
  \bibinfo{journal}{Phys. Rev. B} \textbf{\bibinfo{volume}{65}},
  \bibinfo{pages}{220401(R)} (\bibinfo{year}{2002}).

\bibitem[{\citenamefont{Dauguet et~al.}(1996)\citenamefont{Dauguet, Gandit,
  Chaussy, Lee, Fert, and Holody}}]{Dauguet:prb96}
\bibinfo{author}{\bibfnamefont{P.}~\bibnamefont{Dauguet}},
  \bibinfo{author}{\bibfnamefont{P.}~\bibnamefont{Gandit}},
  \bibinfo{author}{\bibfnamefont{J.}~\bibnamefont{Chaussy}},
  \bibinfo{author}{\bibfnamefont{S.}~\bibnamefont{Lee}},
  \bibinfo{author}{\bibfnamefont{A.}~\bibnamefont{Fert}}, \bibnamefont{and}
  \bibinfo{author}{\bibfnamefont{P.}~\bibnamefont{Holody}},
  \bibinfo{journal}{Phys. Rev. B} \textbf{\bibinfo{volume}{54}},
  \bibinfo{pages}{1083} (\bibinfo{year}{1996}).

\bibitem[{\citenamefont{Vedyayev et~al.}(1997)\citenamefont{Vedyayev,
  Ryzhanova, Dieny, Dauguet, Gandit, and Chaussy}}]{Vedyayev:prb97}
\bibinfo{author}{\bibfnamefont{A.}~\bibnamefont{Vedyayev}},
  \bibinfo{author}{\bibfnamefont{N.}~\bibnamefont{Ryzhanova}},
  \bibinfo{author}{\bibfnamefont{B.}~\bibnamefont{Dieny}},
  \bibinfo{author}{\bibfnamefont{P.}~\bibnamefont{Dauguet}},
  \bibinfo{author}{\bibfnamefont{P.}~\bibnamefont{Gandit}}, \bibnamefont{and}
  \bibinfo{author}{\bibfnamefont{J.}~\bibnamefont{Chaussy}},
  \bibinfo{journal}{Phys. Rev. B} \textbf{\bibinfo{volume}{55}},
  \bibinfo{pages}{3728} (\bibinfo{year}{1997}).

\bibitem[{\citenamefont{Giacomoni et~al.}(unpublished)\citenamefont{Giacomoni,
  Dieny, {Pratt, Jr.}, Loloee, and Tsoi}}]{Giacomoni:02}
\bibinfo{author}{\bibfnamefont{L.}~\bibnamefont{Giacomoni}},
  \bibinfo{author}{\bibfnamefont{B.}~\bibnamefont{Dieny}},
  \bibinfo{author}{\bibfnamefont{W.}~\bibnamefont{{Pratt, Jr.}}},
  \bibinfo{author}{\bibfnamefont{R.}~\bibnamefont{Loloee}}, \bibnamefont{and}
  \bibinfo{author}{\bibfnamefont{M.}~\bibnamefont{Tsoi}}
  (\bibinfo{year}{unpublished}).

\bibitem[{\citenamefont{Bauer et~al.}(2003{\natexlab{a}})\citenamefont{Bauer,
  Tserkovnyak, Huertas-Hernando, and Brataas}}]{Gerrit:prb03}
\bibinfo{author}{\bibfnamefont{G.~E.~W.} \bibnamefont{Bauer}},
  \bibinfo{author}{\bibfnamefont{Y.}~\bibnamefont{Tserkovnyak}},
  \bibinfo{author}{\bibfnamefont{D.}~\bibnamefont{Huertas-Hernando}},
  \bibnamefont{and} \bibinfo{author}{\bibfnamefont{A.}~\bibnamefont{Brataas}},
  \bibinfo{journal}{Phys. Rev. B} \textbf{\bibinfo{volume}{67}},
  \bibinfo{pages}{094421} (\bibinfo{year}{2003}{\natexlab{a}}).

\bibitem[{\citenamefont{Urazhdin
  et~al.}(2004{\natexlab{a}})\citenamefont{Urazhdin, Loloee, and {Pratt,
  Jr.}}}]{Urazhdin:prb05}
\bibinfo{author}{\bibfnamefont{S.}~\bibnamefont{Urazhdin}},
  \bibinfo{author}{\bibfnamefont{R.}~\bibnamefont{Loloee}}, \bibnamefont{and}
  \bibinfo{author}{\bibfnamefont{W.}~\bibnamefont{{Pratt, Jr.}}},
  \bibinfo{journal}{Phys. Rev. B} \textbf{\bibinfo{volume}{71}},
  \bibinfo{pages}{100401(R)} (\bibinfo{year}{2004}{\natexlab{a}}).

\bibitem[{\citenamefont{Kovalev et~al.}(2002)\citenamefont{Kovalev, Brataas,
  and Bauer}}]{Kovalev:prb02}
\bibinfo{author}{\bibfnamefont{A.~A.} \bibnamefont{Kovalev}},
  \bibinfo{author}{\bibfnamefont{A.}~\bibnamefont{Brataas}}, \bibnamefont{and}
  \bibinfo{author}{\bibfnamefont{G.~E.~W.} \bibnamefont{Bauer}},
  \bibinfo{journal}{Phys. Rev. B} \textbf{\bibinfo{volume}{66}},
  \bibinfo{pages}{224424} (\bibinfo{year}{2002}).

\bibitem[{\citenamefont{Manschot
  et~al.}(2004{\natexlab{a}})\citenamefont{Manschot, Brataas, and
  Bauer}}]{Manschot:prb04}
\bibinfo{author}{\bibfnamefont{J.}~\bibnamefont{Manschot}},
  \bibinfo{author}{\bibfnamefont{A.}~\bibnamefont{Brataas}}, \bibnamefont{and}
  \bibinfo{author}{\bibfnamefont{G.~E.~W.} \bibnamefont{Bauer}},
  \bibinfo{journal}{Phys. Rev. B} \textbf{\bibinfo{volume}{69}},
  \bibinfo{pages}{092407} (\bibinfo{year}{2004}{\natexlab{a}}).

\bibitem[{\citenamefont{Manschot
  et~al.}(2004{\natexlab{b}})\citenamefont{Manschot, Brataas, and
  Bauer}}]{Manschot:apl04}
\bibinfo{author}{\bibfnamefont{J.}~\bibnamefont{Manschot}},
  \bibinfo{author}{\bibfnamefont{A.}~\bibnamefont{Brataas}}, \bibnamefont{and}
  \bibinfo{author}{\bibfnamefont{G.}~\bibnamefont{Bauer}},
  \bibinfo{journal}{Appl. Phys. Lett.} \textbf{\bibinfo{volume}{85}},
  \bibinfo{pages}{3250} (\bibinfo{year}{2004}{\natexlab{b}}).

\bibitem[{\citenamefont{Zwierzycki et~al.}(2005)\citenamefont{Zwierzycki,
  Tserkovnyak, Kelly, Brataas, and Bauer}}]{Zwierzycki:prb05}
\bibinfo{author}{\bibfnamefont{M.}~\bibnamefont{Zwierzycki}},
  \bibinfo{author}{\bibfnamefont{Y.}~\bibnamefont{Tserkovnyak}},
  \bibinfo{author}{\bibfnamefont{P.}~\bibnamefont{Kelly}},
  \bibinfo{author}{\bibfnamefont{A.}~\bibnamefont{Brataas}}, \bibnamefont{and}
  \bibinfo{author}{\bibfnamefont{G.~E.~W.}~\bibnamefont{Bauer}},
  \bibinfo{journal}{Phys. Rev. B} \textbf{\bibinfo{volume}{71}},
  \bibinfo{pages}{64420}  (\bibinfo{year}{2005}).

\bibitem[{\citenamefont{Shpiro et~al.}(2003)\citenamefont{Shpiro, Levy, and
  Zhang}}]{Shpiro:prb03}
\bibinfo{author}{\bibfnamefont{A.}~\bibnamefont{Shpiro}},
  \bibinfo{author}{\bibfnamefont{P.~M.} \bibnamefont{Levy}}, \bibnamefont{and}
  \bibinfo{author}{\bibfnamefont{S.}~\bibnamefont{Zhang}},
  \bibinfo{journal}{Phys. Rev. B} \textbf{\bibinfo{volume}{67}},
  \bibinfo{pages}{104430} (\bibinfo{year}{2003}).

\bibitem[{\citenamefont{Berger}(2003)}]{Berger:jap03}
\bibinfo{author}{\bibfnamefont{L.}~\bibnamefont{Berger}}, \bibinfo{journal}{J.
  Appl. Phys.} \textbf{\bibinfo{volume}{93}}, \bibinfo{pages}{7693}
  (\bibinfo{year}{2003}).

\bibitem[{\citenamefont{Nakamura}(2004)}]{Nakamura:matj04}
\bibinfo{author}{\bibfnamefont{S.}~\bibnamefont{Nakamura}},
  \bibinfo{journal}{Materia Japan} \textbf{\bibinfo{volume}{43}},
  \bibinfo{pages}{498} (\bibinfo{year}{2004}).

\bibitem[{\citenamefont{Tserkovnyak et~al.}(2002)\citenamefont{Tserkovnyak,
  Brataas, and Bauer}}]{Tserkovnyak:prl02}
\bibinfo{author}{\bibfnamefont{Y.}~\bibnamefont{Tserkovnyak}},
  \bibinfo{author}{\bibfnamefont{A.}~\bibnamefont{Brataas}}, \bibnamefont{and}
  \bibinfo{author}{\bibfnamefont{G.~E.~W.} \bibnamefont{Bauer}},
  \bibinfo{journal}{Phys. Rev. Lett.} \textbf{\bibinfo{volume}{88}},
  \bibinfo{pages}{117601} (\bibinfo{year}{2002}).

\bibitem[{\citenamefont{Brataas et~al.}(2003)\citenamefont{Brataas, Zarand,
  Tserkovnyak, and Bauer}}]{Brataas:prl03}
\bibinfo{author}{\bibfnamefont{A.}~\bibnamefont{Brataas}},
  \bibinfo{author}{\bibfnamefont{G.}~\bibnamefont{Zarand}},
  \bibinfo{author}{\bibfnamefont{Y.}~\bibnamefont{Tserkovnyak}},
  \bibnamefont{and} \bibinfo{author}{\bibfnamefont{G.~E.~W.}
  \bibnamefont{Bauer}}, \bibinfo{journal}{Phys. Rev. Lett.}
  \textbf{\bibinfo{volume}{91}}, \bibinfo{pages}{166601}
  (\bibinfo{year}{2003}).

\bibitem[{\citenamefont{Xiao et~al.}(2004)\citenamefont{Xiao, Zangwill, and
  Stiles}}]{Xiao:prb04}
\bibinfo{author}{\bibfnamefont{J.}~\bibnamefont{Xiao}},
  \bibinfo{author}{\bibfnamefont{A.}~\bibnamefont{Zangwill}}, \bibnamefont{and}
  \bibinfo{author}{\bibfnamefont{M.}~\bibnamefont{Stiles}},
  \bibinfo{journal}{Phys. Rev. B} \textbf{\bibinfo{volume}{70}},
  \bibinfo{pages}{172405} (\bibinfo{year}{2004}).

\bibitem[{\citenamefont{{Pratt, Jr.} and et~al}(1997)}]{Pratt:ieeem97}
\bibinfo{author}{\bibfnamefont{W.}~\bibnamefont{{Pratt, Jr.}}}
  \bibnamefont{and} \bibinfo{author}{\bibnamefont{et~al}},
  \bibinfo{journal}{IEEE Trans. Mag.} \textbf{\bibinfo{volume}{33}},
  \bibinfo{pages}{3505} (\bibinfo{year}{1997}).

\bibitem[{\citenamefont{Urazhdin
  et~al.}(2004{\natexlab{b}})\citenamefont{Urazhdin, Loloee, and {Pratt,
  Jr.}}}]{Urazhdin:cm04}
\bibinfo{author}{\bibfnamefont{S.}~\bibnamefont{Urazhdin}},
  \bibinfo{author}{\bibfnamefont{R.}~\bibnamefont{Loloee}}, \bibnamefont{and}
  \bibinfo{author}{\bibfnamefont{W.}~\bibnamefont{{Pratt, Jr.}}}
  (\bibinfo{year}{2004}{\natexlab{b}}), \bibinfo{note}{cond-mat/0403441 v1}.

\bibitem[{\citenamefont{Zhang et~al.}(2002)\citenamefont{Zhang, Levy, and
  Fert}}]{Zhang:prl02}
\bibinfo{author}{\bibfnamefont{S.}~\bibnamefont{Zhang}},
  \bibinfo{author}{\bibfnamefont{P.~M.} \bibnamefont{Levy}}, \bibnamefont{and}
  \bibinfo{author}{\bibfnamefont{A.}~\bibnamefont{Fert}},
  \bibinfo{journal}{Phys. Rev. Lett.} \textbf{\bibinfo{volume}{88}},
  \bibinfo{pages}{236601} (\bibinfo{year}{2002}).

\bibitem[{\citenamefont{Bauer et~al.}(2003{\natexlab{b}})\citenamefont{Bauer,
  Brataas, Tserkovnyak, and van Wees}}]{Bauer:apl03}
\bibinfo{author}{\bibfnamefont{G.~E.~W.} \bibnamefont{Bauer}},
  \bibinfo{author}{\bibfnamefont{A.}~\bibnamefont{Brataas}},
  \bibinfo{author}{\bibfnamefont{Y.}~\bibnamefont{Tserkovnyak}},
  \bibnamefont{and} \bibinfo{author}{\bibfnamefont{B.~J.} \bibnamefont{van
  Wees}}, \bibinfo{journal}{Appl. Phys. Lett.} \textbf{\bibinfo{volume}{82}},
  \bibinfo{pages}{3928} (\bibinfo{year}{2003}{\natexlab{b}}).

\end{thebibliography}

\end{document}